\documentclass[aps,prl,twocolumn,showpacs,superscriptaddress,longbibliography]{revtex4-1}
\pdfoutput=1

\usepackage{graphicx}
\usepackage{dcolumn}
\usepackage{color}

\usepackage{eucal}
\usepackage{amssymb}

\usepackage{amsmath,amsthm}

\begin{document}
\title{Travelling fronts in active-passive particle mixtures}

\author{Adam Wysocki, Roland G. Winkler, Gerhard Gompper}
\affiliation{Theoretical Soft Matter and Biophysics, 
Institute of Complex Systems and Institute for Advanced Simulation,
Forschungszentrum J\"ulich, 52425 J\"ulich, Germany}

\date{\today}
\begin{abstract}
The emergent dynamics in phase-separated mixtures of isometric active and
passive Brownian particles is studied numerically in two dimensions.
A novel steady-state of well-defined traveling fronts is observed, where the interface
between the dense and the dilute phase propagates and the bulk of both phases
is (nearly) at rest. Two kind of interfaces, advancing and receding, are 
formed by spontaneous symmtry breaking, induced by an instability of a planar interface due to the 
formation of localized vortices. The propagation arises due to flux imbalance at the interface, 
strongly resembling travelling fronts in reaction-diffusion systems. 
Above a threshold, the interface velocity decreases linearly with
increasing fraction of active particles. 
\end{abstract}

\pacs{82.70.Dd,64.75.Xc}



\maketitle

{\it Introduction} -- 
Generic models of active fluids divide into two main classes,
systems of active Brownian particles (discs or spheres) \cite{elgeti2015rop,marchetti2015arxiv},
which emphasize volume exclusion,
and systems of anisotropic (or elongated) self-propelled particles, which emphasize alignment
interactions, like the Vicsek model \cite{gregoire2004prl,solon2015prl}.
The most striking phenomenon of active Brownian particles (ABPs), observed in various experiments
\cite{theurkauff2012prl,palacci2013science,buttinoni2013prl,ginot2015prx} and simulations
\cite{fily2014sm1,redner2013prl,myself2014epl},
is motility-induced phase separation \cite{cates2015arcmp}. The phase behavior
and kinetics, like domain coarsening \cite{redner2013prl,stenhammar2014sm} or interface
fluctuations \cite{bialke2015prl}, of ABPs resemble a passive fluid with attractive interaction. 
Moreover, ABPs exhibit an intriguing collective dynamics with jets 
and swirls, which has been speculated to arise from interfacial 
sorting of ABPs with different orientations \cite{myself2014epl}.
In contrast, the kinetics of models with alignment interaction exhibits various
modes of collective motion, for example large polar swarms, i.e., high-density polar bands travelling
coherently through an isotropic background gas \cite{gregoire2004prl,solon2015prl}.

A natural extension of single-component ABP fluids are mixtures of particles with, e.g.,
different activities \cite{ni2014sm,stenhammar2015prl,kummel2015sm,takatori2015sm}, 
temperatures \cite{grosberg2015pre,weber2015arxiv}, or diameters \cite{yang2014sm}. 
These models also exhibit activity-induced phase separation. 
We focus here on the dynamics of mixtures of isometric active and passive Brownian particles.
Surprisingly, we find that these systems exhibit a novel and so far unexplored
type of collective motion in the phase-separated state in the form of well-defined propagating fronts, 
which can be either enriched or depleted of active particles, and are advancing toward or 
receding from the dense phase, respectively. The propagation arises due to flux imbalance 
at the interface between the dense and dilute phases with a strong resemblance to travelling fronts 
in reaction-diffusion systems \cite{tyson1988phyD}. The selection of the interface type 
(advancing or receding) happens by spontaneous symmetry breaking, 
induced by an instability of a planar interface due to the formation of localized vortices.
In contrast to the polar bands of the Vicsek model \cite{gregoire2004prl,solon2015prl}, 
which travel as a whole, here only the interface between the two phases propagates.

The recent finding of a stable interface in pure, phase-separated ABP fluids, together with
a negative surface tension \cite{bialke2015prl}, questions the mapping of an ABP fluid onto an 
equilibrium fluid with attraction. The emergence of travelling fronts in active-passive 
mixtures clearly contradicts the existence of an equivalent equilibrium system in this case. 
Instead, we propose an explanation of the interface dynamics on the basis of well-studied 
models describing the growth of rough surfaces under far-from-equilibrium conditions 
\cite{barabasi1995,krug1997ainp}.

\begin{figure}[!]
\centering
\includegraphics[width=1\columnwidth]{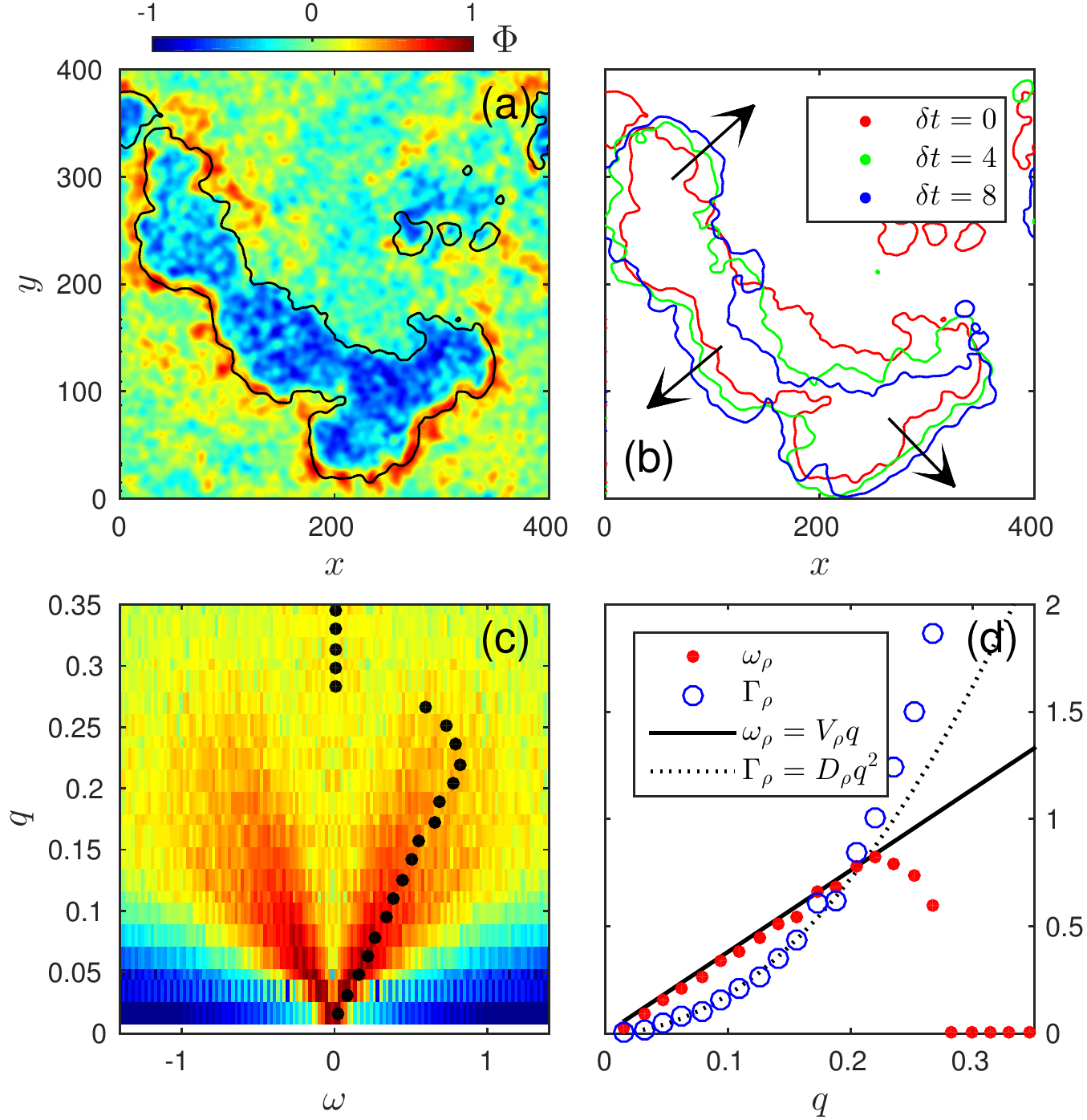}
\caption{\label{f:snapshots_dyn_struct}(color online)
(a) Snapshot of the segregation order parameter field $\Phi(\mathbf{r})$ of a system of
size $L_x=L_y=400$ with $N=160000$ particles ($\phi=0.78$) in the steady state at $x_A=0.5$. 
$\Phi=1$ ($\Phi=-1$) corresponds to a pure active (passive) phase and 
$\Phi=0$ corresponds to a uniform mixture. The black line indicates the interface position.
(b) Time evolution of the interface position, see movies in \cite{SI}.
(c) Dynamic structure factor $S(\omega,q)$ as a function of frequency $\omega$ and wavenumber $q$.
The shape of $S(\omega,q)$ indicates a damped propagative mode.
(d) Frequency $\omega_{\rho}$ (also indicated in (c) by black dots) and
damping rate $\Gamma_{\rho}$ as function of wavenumber. Lines indicate
$\omega_{\rho}=V_{\rho}q$, with $V_{\rho}\approx0.04V_0$, and
$\Gamma_{\rho}=D_{\rho}q^2$, with $D_{\rho}\approx0.004D_0$, where $D_0$ is
the diffusion constant of a free ABP.}
\end{figure}

{\it Model} -- 
We simulate a mixture of $N_A$ active and $N_P$ passive Brownian disks
(in total $N=N_A+N_P$ particles) in a 2D simulation box of size $L_x\times L_y$
with periodic boundary conditions. Their dynamics is overdamped, i.e.,
$\dot{\mathbf{r}}_i=V_0\mathbf{e}_i+\mathbf{f}_i/\gamma_t+\boldsymbol{\xi}_i$,
where $V_0$ is the propulsion velocity along the polar axis $\mathbf{e}_i$
($V_0=0$ for passive particles), $\mathbf{f}_i=\sum_{i\neq j}\mathbf{f}_{ij}$ is
the force due to steric interactions, and $\boldsymbol{\xi}_i$ is a random velocity.
The particles interact via the soft repulsive force
$\mathbf{f}_{ij}=k(a_i+a_j-|\mathbf{r}_{ij}|)\mathbf{r}_{ij}/|\mathbf{r}_{ij}|$,
with $\mathbf{r}_{ij}=\mathbf{r}_i-\mathbf{r}_i$ if $a_i+a_j<|\mathbf{r}_{ij}|$
and $\mathbf{f}_{ij}=0$ otherwise \cite{fily2014sm1}. The discs are polydisperse in order
to avoid crystallization and their radii $a_i$ are uniformly distributed
in the interval $[0.8a, 1.2a]$ \cite{fily2014sm1}. 
The zero-mean Gaussian white-noise velocity $\boldsymbol{\xi}_i$ obeys
$\langle\boldsymbol{\xi}_i(t)\boldsymbol{\xi}_i(t')\rangle=2D_t\delta_{ij}\boldsymbol{1}\delta(t-t')$,
where $D_t=k_BT/\gamma_t$ is the translational diffusion coefficient with thermal energy $k_BT$ 
and friction coefficient $\gamma_t$. The propulsion direction $\mathbf{e}_i$ undergoes
a free rotational diffusion with a diffusion constant $D_r$,
where $D_t/D_r=4a^2/3$ holds for a no-slip sphere.
The persistence of swimming is characterised by the P\'eclet number $Pe=V_0/(2a D_r)$.
The typical particle overlap due to activity, $\gamma_tV_0/(2a k)$, is fixed to 0.01 \cite{stenhammar2014sm}. 
Unless otherwise noted, we consider systems with $Pe=100$ and an packing fraction 
$\phi=\sum_{i=1}^N\pi a_i^2/(L_xL_y)=0.67$, below random closed packing \cite{fily2014sm1},
and vary the fraction $x_A=N_A/N$ of ABPs. Moreover, lengths are expressed in units of $2a$
and time in units of $1/D_r$.

{\it Phase behavior} -- 
A mixture of active and passive discs separates into a dense and a dilute phase at sufficiently 
large $Pe$, $\phi$, and $x_A$ \cite{stenhammar2015prl}, very similar to a pure ABP fluid \cite{redner2013prl}; 
in addition, active and passive particles tend to segregate inside the dense phase. This is illustrated in 
Fig.~\ref{f:snapshots_dyn_struct}(a), where the segregation order parameter
field $\Phi(\mathbf{r})$ of a large phase-separated system with curved interfaces between the
dense and the dilute phase is shown. $\Phi$ is defined as
$\Phi(\mathbf{r})=(\phi_A-\phi_P)/(\phi_A+\phi_P)$ with coarse-grained packing density fields
$\phi_A(\mathbf{r})$ and $\phi_P(\mathbf{r})$ of active and passive particles, respectively \cite{SI}. 
The dilute phase consist mainly of passive particles ($\Phi\approx-1$) and the bulk of the
dense phase is a homogenous active-passive mixture ($\Phi\approx0$) with small patches of
enriched active or passive particles. Within the interface region, we observe either an accumulation
($\Phi\approx1$) or a depletion ($\Phi<0$) of active particles. 

Note that a completely different behavior appears for dilute solutions (not considered here), namely, 
for mean free paths much larger than the persistence length of swimming ($\pi\sigma/4\phi\gg V_0/D_r$), 
where active (passive) particles behave as effectively 'hot' ('cold') particles. 
Such a mixture exhibits phase separation into a solid-like cluster of passive and 
a gaseous phase of the active particles \cite{weber2015arxiv}.

{\it Bulk travelling fronts} -- The focus of our paper is on the kinetics of a phase-separated 
active-passive mixture. The domain dynamics of a one-component ABP fluid in the steady state 
is limited to fluctuating interfaces resembling thermal capillary waves \cite{bialke2015prl,SI},
except for the coarsening kinetics \cite{redner2013prl,stenhammar2014sm}.
By contrast, active-passive mixtures exhibit amazingly mobile
or travelling interfaces in a large region of the $Pe-\phi-x_A$ parameter space.
The interface propagation becomes apparent from Fig.~\ref{f:snapshots_dyn_struct}(b),
where the time evolution of the interface position is shown, 
see also movies in Ref.~\cite{SI}.

In order to quantify our observations, we analyse the density correlations at $x_A=0.5$ and $\phi=0.78$ 
by the dynamic structure factor
$S(\mathbf{q},\omega)=\int_{-\infty}^{\infty} F(\mathbf{q},t)\exp{(i\omega t)}\,\mathrm{d}t$,
where $\mathbf{q}$ is the wavevector, $\omega$ is the angular frequency, and
$F(\mathbf{q},t)=\langle\rho_{\mathbf{q}}(t)\rho_{-\mathbf{q}}(0)\rangle/N$
is the correlation function of the Fourier components $\rho_{\mathbf{q}}$ of the density \cite{hansen1990}.
The circularly averaged $S(\mathbf{q},\omega)$, which is accessible by scattering
experiments, is shown in Fig.~\ref{f:snapshots_dyn_struct}(c). 
The structure factor exhibits peaks at the frequencies $\pm \omega_{\rho}$, with
a width, which increases with increasing $q$. This suggests a damped propagative mode \cite{hansen1990} 
related to the traveling interfaces indicated in Fig.~\ref{f:snapshots_dyn_struct}(b). We obtain the full 
dispersion relations $\omega_{\rho}(q)$ and $\Gamma_{\rho}(q)$ by fitting 
$F(q,0)\exp{(-\Gamma_{\rho}t)}\cos{(\omega_{\rho}t)}$ to the corresponding simulation data. 
As can be seen in Fig.~\ref{f:snapshots_dyn_struct}(d),
the peak positions follow the Brillouin-like dispersion relation $\omega_{\rho}=V_{\rho}q$ up to $q\approx 0.2$
with a velocity $V_{\rho}\approx0.04V_0$.
In a simple equilibrium fluid (Newtonian dynamics) such a velocity is the speed of sound \cite{hansen1990}, 
but here, $V_{\rho}$ is related to the velocity of interface propagation. The decay rate $\Gamma_{\rho}(q)$ 
obeys $\Gamma_{\rho}=D_{\rho}q^2$ up to $q\approx 0.2$, with the transport coefficient $D_{\rho}\approx0.004D_0$, 
where $D_0=D_t+V_0^2/(2D_r)$ is the diffusion constant of a free ABP. 
All modes are strongly damped for $q>0.25$, i.e, dense (or dilute) phase droplets with size smaller 
than $50a$ dissolve quickly.

\begin{figure*}[!]
\centering
\includegraphics[width=2\columnwidth]{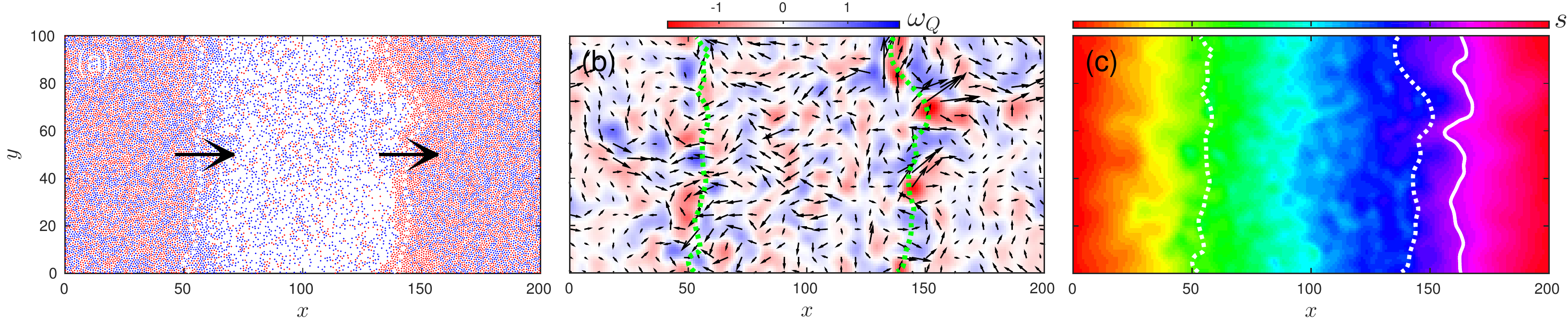}
\caption{\label{f:snapshots_elong}(color online) (a) Snapshot of a mixture of active (red) and
passive (blue) Brownian discs at $x_A=0.5$ in a box of size $L_x=L_y/2=100$.
The interfaces between the two phases (dashed lines) travel by chance to the right . 
(b) Vorticity $\omega_Q(\mathbf{r})=\nabla\times\mathbf{Q}(\mathbf{r})$, where $\mathbf{Q}$ 
is the coarse-grained particle flux \cite{SI}. (c) Visualisation of the bulk flow. 
Particle positions are shown after the time lag $\delta t=2$, 
where particles are colored according to their initial $x$ position, as indicated by the color scale. 
The solid line  marks the isoline of $s(\mathbf{r},\delta t)$ used for the stability analysis. 
See corresponding movies in Ref.~\cite{SI}.}
\end{figure*}
\begin{figure}[!]
\centering
\includegraphics[width=1\columnwidth]{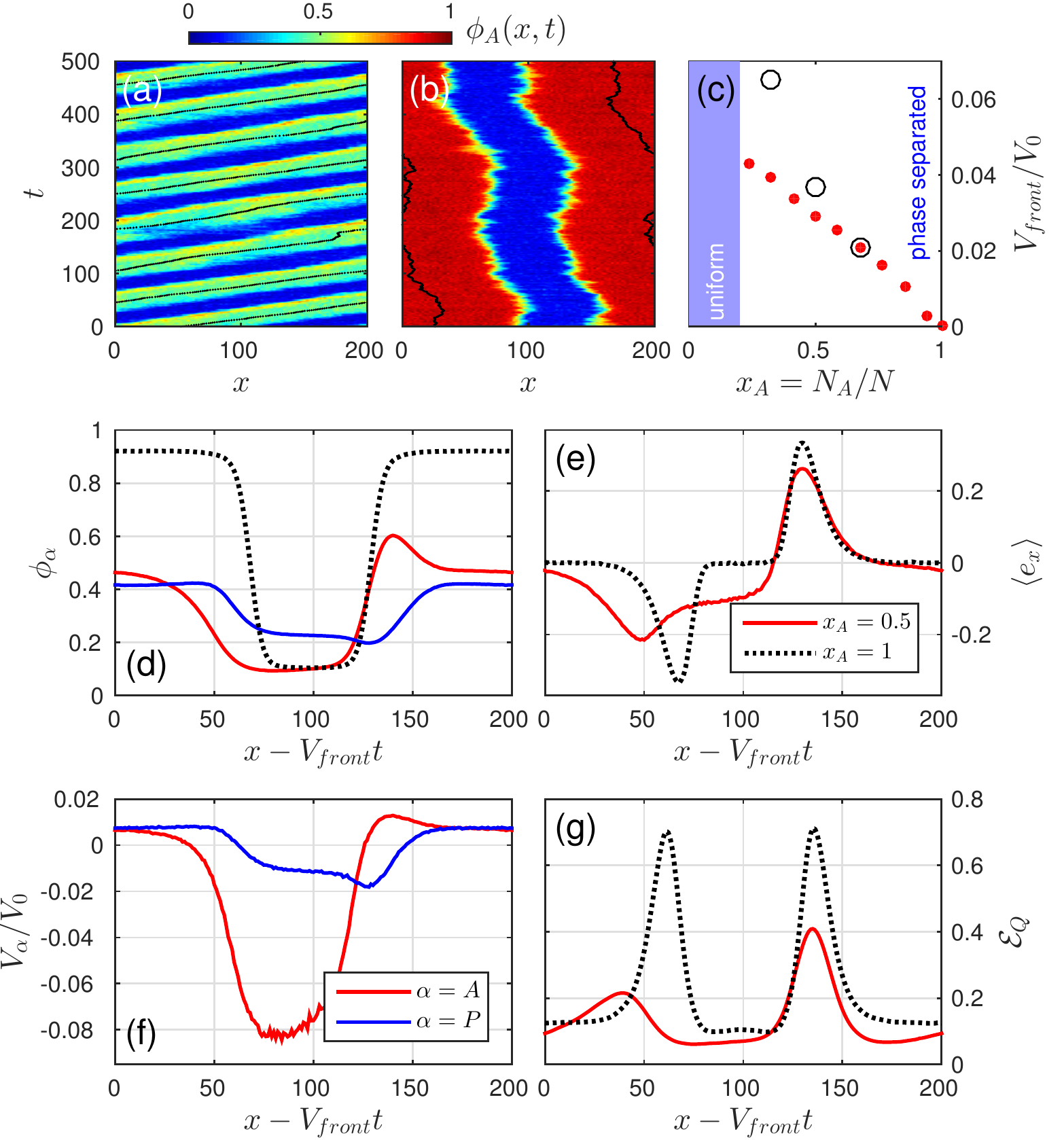}
\caption{\label{f:mean_profile}(color online) Space-time dependence of the 
active-particle packing density $\phi_A(x,t)$ for (a) the mixture with 
$x_A=0.5$ and (b) a pure active fluid with $x_A=1$. The
black lines indicate the center-of-mass position of $\phi_A(x,t)$.
(c) Velocities $V_{front}$ (bullets) of the interface and $V_{\rho}$ (circles)
extracted from $S(q,\omega)$, see Fig.~\ref{f:snapshots_dyn_struct}, 
as function of $x_A$. Phase separation appears for $x_A=0.2$. 
(d-g) Time averaged profiles measured in a comoving frame, i.e., relative to the propagating front, 
(d) of active- and passive-particle packing fractions $\phi_A(x)$ and $\phi_P(x)$, 
(e) of the active-particle $x$-polaristion $\langle e_x \rangle(x)$, 
(f) of active- and passive-particle velocities along the front propagation direction $V_A(x)$ and $V_P(x)$, 
and (g) of the intensity of the particle flux vorticity $\mathcal{E}_Q(x)$. 
We choose a representation, where the front propagates to the right. 
Solid lines correspond to $x_A=0.5$ and dashed lines to $x_A=1$.}
\end{figure}

{\it Stabilized travelling fronts} -- In order to study a propagating interface
in more detail, we employ a quasi-one-dimensional setup of an elongated box of lengths $L_x=2L_y$ 
\cite{vink2005jcp,bialke2015prl} such that the interface favors to span the shorter box length, 
see Fig.~\ref{f:snapshots_elong}(a). Given that the system phase-separates, such a configuration forms 
spontaneously and remains stable over long time. In Fig.~\ref{f:mean_profile} (a,b), we show the time evolution 
of the active-particle packing profile $\phi_A(x,t)$ averaged over the $y$-coordinate.
The interface position in a pure active fluid performs a diffusive motion \cite{SI}. 
By contrast, in our mixture the translational symmetry is broken and both interfaces propagate steadily
in parallel (for the set up of Fig.~\ref{f:snapshots_elong}) either to the right or to the left 
with equal probability \cite{SI}. The travelling front is extremely stable within the typical simulation 
length of $T=5\times10^3$, however, the steady propagation is occasionally interrupted by 
intermittent large-scale rearrangements of the front. The front velocity $V_{front}$ is nearly 
independent of the overall packing fraction $\phi$ and activity $Pe$. However, $V_{front}$ 
monotonously decreases with increasing $x_A$ 
[in a good agreement with $V_{\rho}$ obtained from $S(q,\omega)$], with a maximum just above 
the active-particle fraction, where phase separation sets in, see Fig.~\ref{f:mean_profile} (c).

We calculate profiles of various quantities in a comoving frame. 
In a pure active fluid, the two interfaces are equivalent and the dense phase is symmetrically
surrounded by a corona of particles with their polar vector pointing preferentially toward that phase, see dashed lines in
Fig.~\ref{f:mean_profile}(d,e). In a mixture, this symmetry is broken and active particles
preferentially accumulate at one interface and deplete at the other. Similarly, the $x$-polaristion
$\langle e_x \rangle$ at both interfaces is different; it is larger at the side of preferred accumulation.
Moreover, $\langle e_x \rangle$ takes negative values in the dilute phase, accompanied by
a negative $x$-velocity of active particles, $V_A$, causing in turn a negative
velocity of passive particles, $V_P$, due to collisions between passive and active particles,
see Fig.~\ref{f:mean_profile}(f). 
The mass transport from the dense into the dilute phase is characterized by the intensity 
of the particle flux vorticity $\mathcal{E}_Q(x)=\int_{0}^{L_y}\omega_Q^2(\mathbf{r})\,\mathrm{d}y/L_y$ 
(see detailed discussion below), where $\omega_Q(\mathbf{r})$ is the curl of the particle flux 
$\mathbf{Q}(\mathbf{r})$. $\mathcal{E}_Q$ is most pronounced within the interface region and, 
in case of a mixture, $\mathcal{E}_Q$ is larger at the side of larger polarisation and 
active particle accumulation, see Fig.~\ref{f:mean_profile}(g) and Fig.~\ref{f:snapshots_elong}(b).

\begin{figure*}[t]
\centering
\includegraphics[width=1.9\columnwidth]{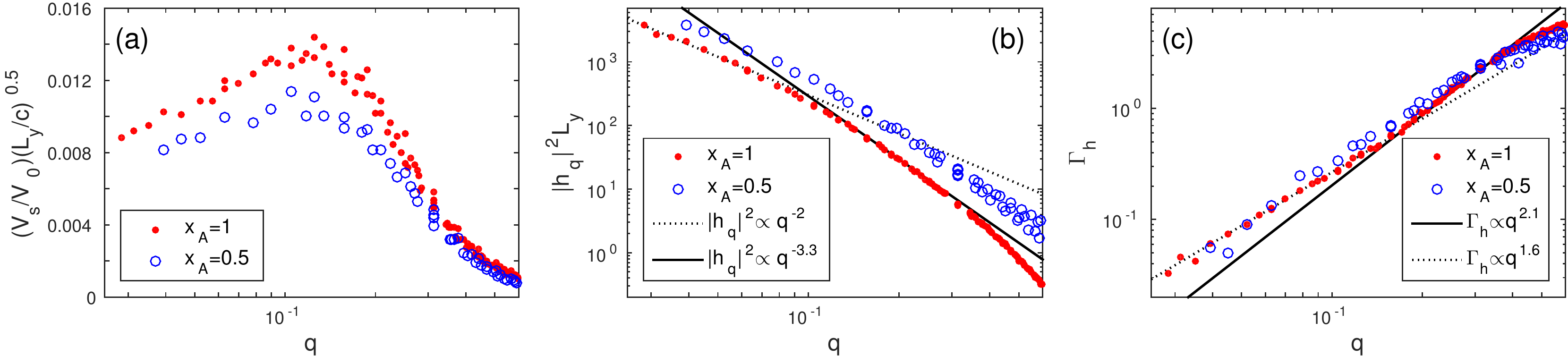}
\caption{\label{f:spectrum}(color online) (a) Growth velocities $V_s(q)$ of the Fourier modes 
of an isoline of the scalar displacement field $s(\mathbf{r},\delta t)$ inside the dense phase,
see Fig.~\ref{f:snapshots_elong}(c). 
(b) Spectrum of the interfacial height fluctuations $\langle|h_q|^2\rangle$ as a function of 
the wave number $q$; results for different $L_y$ at fixed $L_y/L_x$ are shown. 
(c) Decay rate $\Gamma_h(q)$ of the interfacial height autocorrelation function 
$\langle h_q(t)h_{-q}(0)\rangle$. Different scaling regimes are indicated by lines.}
\end{figure*}
 
{\it Discussion of interfaces in pure ABP fluids} --
Interfaces in pure ABP fluids are not propagating, but show a diffusive dynamics as evident from 
Fig.~\ref{f:mean_profile}(b) and the movies in Ref.~\cite{SI}. Within the picture of motility-induced
phase separation, there is a balance between an active flux of particles from
the low-density gas phase colliding with the dense phase
and a diffusive flux of particles leaving the high-density phase due to
rotational diffusion \cite{redner2013prl}.
This alone would generate a very rough and uncorrelated interface
structure, like in random particle deposition \cite{barabasi1995}.

However, activity also leads to smoothing of the interface.  
For an undulated interface, particles which bump into the interface
slide into regions of high convexity \cite{fily2014sm2,myself2014epl}, thereby level out 
the interface on small scales and produce a local polarisation. In turn, this local
polarisation induces an internal mass flow inside the dense phase
such that alternating vortices of opposite vorticity emerge within the
interface region due to mass conservation, see
Fig.~\ref{f:mean_profile}(g) and movies in Ref.~\cite{SI}.
As a result, randomly oriented particles emerge from the bulk and cause
an evaporation of the interface protrusions. 

{\it Discussion of interfaces in active-passive mixtures} --
This picture changes considerably in mixtures, where the interfaces are
propagating due to the dynamical coupling of active and passive particles. 
Imagine a perturbation such that the polarisation in the right interface region of
Fig.~\ref{f:snapshots_elong}(a) is larger then that in the
left one. Hence, more active particles leave the right interface, due
to a larger vorticity, and the collision-induced flux of passive particles from right to left exceeds the opposite flux.
Now, a corona of passive particles covers the dense phase in the left interface region and
inhibits the further discharge of active particles. This is supported by
the shape of the profiles
$\phi_A(x)$, $\phi_P(x)$, $\langle e_x \rangle(x)$, and $\mathcal{E}_Q(x)$ in Fig.~\ref{f:mean_profile}.
The overall picture that the traveling front is a consequence of a flux imbalance and
that the particle transport out of the interface is dominated by vortex formation
and not by rotational diffusion is in line with the fact that $V_{front}$ is independent
of $Pe$, if $V_0$ is fixed and $D_r$ is varied via $k_BT$. 
In order to quantify the vorticity-induced mass transport, we use the horizontal 
positions $x_i(t)$ of all particles $i$ (at time $t$) to construct a scalar displacement 
field $s(\mathbf{r},\delta t)$ \cite{SI}. The temporal evolution 
of $s(\mathbf{r},\delta t)$ then indicates the particle convection, see 
Fig.~\ref{f:snapshots_elong}(c) and movies in Ref.~\cite{SI}. 
We choose a isoline of $s(\mathbf{r},\delta t)$ 
inside the dense phase; this isoline is initially flat, but roughens as a function of time.
This process is monitored by the Fourier modes of this isoline, 
similar to a stability analysis of the Rayleigh-Taylor instability. 
The Fourier-mode amplitudes grow with constant velocity $V_s(q)$ at short times. 
The growth velocities first increase with increasing $q$, reach a maximum at $q \approx 0.1$, 
and exhibit a fast decay for larger $q$, see Fig.~\ref{f:spectrum}(a). 
This confirms the visual impression in Fig.~\ref{f:snapshots_elong}(b,c)
of a characteristic length scale of internal mass currents.

{\it Interface correlations and relaxation} -- 
We analyse the structure and dynamics of the interface by a Fourier transform of its fluctuations. 
From the Fourier amplitudes $h_q$ \cite{SI}, we obtain the interface structure factor 
$S(q)=\langle|h_q|^2\rangle$ and the autocorrelation function $\langle h_q(t)h_{-q}(0)\rangle$, 
which we fit by $S(q)\exp{(-\Gamma_{h}t)}$ to obtain the damping rate $\Gamma_h(q)$ as function of $q$.
We observe a length-scale-dependent scaling $S\propto q^{-(1+2\alpha)}$ and $\Gamma_h\propto q^z$,
where $\alpha$ and $z$ are the roughness and the dynamic exponent, respectively \cite{siegert1996pre}.
We find $\alpha\approx1/2$ and $z\approx1.6$ on large scales, $q\lesssim0.1$, and
$\alpha\approx1$ and $z\approx2$ on intermediate scales, $0.1\lesssim q\lesssim0.4$, for
static as well as traveling interfaces, see Fig.~\ref{f:spectrum}(b,c). In comparison,
an overdamped fluid interface with thermally excited capillary waves in equilibrium
has $\alpha=1/2$ \cite{vink2005jcp} and $z=1$ \cite{gross2013pre} for $q<0.6$.
However, physically more related is the Edward-Wilkinson model \cite{barabasi1995,krug1997ainp},
for non-equilibrium interface growth -- where random particle arrival leads to interface roughening, 
while lateral motion (e.g., due to gravity) yields interface smoothing -- with exponents 
$\alpha=1/2$ and $z=2$. If additionally local growth perpendicular to the interface is present, as
in the Kardar-Parisi-Zhang model, the dynamic exponent $z=3/2$ is expected \cite{barabasi1995,krug1997ainp},
very close to the exponents characterizing the interface behavior of our active particle fluids.

{\it Conclusions} -- 
Active-particle systems display many unexpected features -- both static and dynamic.
We have shown that the large-scale interface structure in mixtures is similar to that 
at equilibrium, however, the dynamics, like interface relaxation or front propagation, 
exhibits strong nonequilibrium characteristics. 
Our results call for an experimental investigation over a wide range of concentrations
and activities. Active-passive mixtures could be realized experimentally by active colloids \cite{theurkauff2012prl,palacci2013science,buttinoni2013prl,ginot2015prx}, 
vibrated polar disks \cite{deseigne2010prl} or even robots \cite{rubenstein2014science,mijalkov2015arxiv}.

\acknowledgements
We thank A. Varghese and J. Horbach for helpful discussions. The support
by the DFG priority program SPP1726 on ``Microswimmers" is gratefully acknowledged.


%

\end{document}